\documentstyle[prl,epsf,aps]{revtex}

\newcommand \be {\begin{equation}}
\newcommand \bea {\begin{eqnarray}}
\newcommand \ee {\end{equation}}
\newcommand \eea {\end{eqnarray}}
 \newcommand \eps {\epsilon}
 \newcommand \bi {\bibitem}
\newcommand \s {\sigma}

\makeatletter
\newcommand\erfc{\mathop{\operator@font erfc}\nolimits}
\makeatother
\input{epsf}
\begin{document}
\twocolumn[\hsize\textwidth\columnwidth\hsize\csname@twocolumnfalse\endcsname
\draft       

\title{Damage spreading in the mode-coupling equations for glasses}
\author{M. Heerema and F. Ritort} 
\address{
(*) Institute of Theoretical Physics\\ University of Amsterdam\\
Valckenierstraat 65\\ 1018 XE Amsterdam (The Netherlands).\\
E-Mail: heerema@phys.uva.nl,ritort@phys.uva.nl}

\date{\today}
\maketitle

\begin{abstract}
We examine the problem of damage spreading in the off-equilibrium mode
coupling equations. The study is done for the spherical $p$-spin model
introduced by Crisanti, Horner and Sommers. For $p>2$ we show the
existence of a temperature transition $T_0$ well above any relevant
thermodynamic transition temperature. Above $T_0$ the asymptotic
damage decays to zero while below $T_0$ it decays to a finite value
independent of the initial damage. This transition is stable in the
presence of asymmetry in the interactions. We discuss the physical
origin of this peculiar phase transition which occurs as a
consequence of the non-linear coupling between the damage and the
two-time correlation functions.
\end{abstract} 

\vfill

\vfill

\twocolumn
\vskip.5pc]

\narrowtext
The theoretical understanding of the dynamical behavior of glasses is
a long outstanding problem in statistical physics which has recently
revealed new aspects related to the underlying mechanism responsible
of the glass transition \cite{SITGES,ANGELL}. While there are still
some obscure points in the theory (i.e. the inclusion of finite time
activated process beyond the mean-field limit) a scenario has emerged
which unifies the dynamical approach (mode-coupling theory) with the
thermodynamic Adam-Gibbs-Di Marzio approach. 
The scenario
for the dynamical behavior of glasses can be summarized in three
different temperatures which separate three different regimes. In the
high-temperature regime $T>T_d$ the system behaves as a liquid and is
very well described by the mode-coupling equations of G\"otze in the
equilibrium regime \cite{GOTZE}. A crossover takes place at $T_d$
where there is a dynamical singularity and the correlation functions
do not decay to zero in the infinite time limit (ergodicity
breaking). This dynamical singularity is a genuine mean-field effect
which turns out to be a crossover temperature when activated processes
are taken into account.  Below $T_d$ the relaxation time (or
viscosity) starts to grow dramatically fast and seems to diverge at
$T_s$ where the configurational entropy vanishes and a thermodynamic
phase transition takes place. The glass transition $T_g$ (as defined
where the viscosity is $10^{13}$ Poise) lies between $T_s$ and $T_d$
and depends on the cooling rate. Hence $T_g$ does not correspond to a
true dynamical singularity. Furthermore, there is a first order
phase transition $T_M$ where the liquid (if cooled sufficiently slow)
crystallizes. 

The essentials of this scenario have been corroborated in the context
of mean-field spin glasses, and in particular in those models with a
one step replica symmetry breaking transition \cite{KITH}. While the
first-order transition temperature $T_M$ is absent in spin glasses
(disorder prevents the existence of a crystal state) the other two
transitions ($T_s,T_d$) have been clearly identified.

The purpose of this paper is the study of the damage spreading in
models for structural glasses. Damage spreading is the study of the
time propagation of a perturbation or damage in the initial condition
of a system. The propagation of the initial damage is a dynamical
effect which has deserved considerably attention in the past
(specially in the context of dynamical systems, for instance, networks
of boolean automata \cite{KA}) because it allows to explore the
structure of the phase space of the system. To investigate damage
spreading a suitable Hamming distance in the space of configurations
is defined. Then we consider two random initial configurations
$\lbrace \s_i,\tau_i\rbrace$ with a given initial distance $D_0$ for
two identical systems which evolve under identical noise realizations
and compute the distance $D(t)$ as a function of time. Eventually one
is interested in the asymptotic long-time behavior of the distance
$D(t)$, i.e $D_{\infty}=\lim_{t\to\infty} D(t)$. In general, three
different regimes can be distinguished. A high-temperature regime
$T>T_{0}$ where $D_{\infty}=0$ independently of the initial distance
$D_0$. A intermediate regime $T_1<T<T_0$ where
$D_{\infty}=D_{\infty}(T)$ is not zero but independent of the initial
distance. And finally a low-temperature regime $T<T_1$ where
$D_{\infty}=D_{\infty}(T,D_0)$ depends on both temperature and initial
distance. While it is widely believed that $T_1$ corresponds to a
thermodynamic phase transition it is not clear what the physical
meaning of $T_0$ is.


In this work we address the problem of damage spreading in the
off-equilibrium mode-coupling equations. We show the existence of the
temperature $T_0$ in glasses well above $T_d$ and $T_s$ in the high
temperature phase. We show that this new transition is a consequence of
the non-linear coupling between the damage and the corresponding
two-times correlation function. This effect is an essential ingredient of
the mode-coupling equations and should be generally valid even beyond
the mean-field limit.  We believe the appearance of this damage
transition is a quite general result in glassy models (with and without
disorder) where the scenario of G\"otze for mode-coupling transitions is
valid.

The simplest solvable model which yields the off-equilibrium mode
coupling equations is the spherical $p$-spin glass model \cite{CHS}. In
this case, the configurations are described by $N$ continuous spin
variables $\lbrace\s_i; 1\le i \le N\rbrace$ which satisfy the spherical
global constraint $\sum_{i=1}^N\s_i^2=N$. The Langevin dynamics of the
model is given by,

\be
\frac{\partial \s_i}{\partial t}=F_i(\lbrace\s\rbrace)-\mu\s_i+\eta_i
\label{eq1}
\ee

where $F_i$ is the force acting on the spin $\s_i$ due to the
interaction with the rest of the spins,

\be
F_i=-\frac{\partial {\cal H}_i}{\partial \s_i}=\frac{1}{(p-1)!}
\sum_{(i_2,i_3,...,i_p)}\,J_i^{i_2,i_3,..,i_p}\s_{i_2}\s_{i_3}..\s_{i_p}
\label{eq2}
\ee

and ${\cal H}$ is a Hamiltonian. The $J_i^{i_2,i_3,..,i_p}$ are
quenched random variables with zero mean and variance $p!/(2N^{p-1})$
which we take to be symmetric under permutation of the different super
indices.  The calculations presented here can be easily generalized to
asymmetric couplings \cite{CKLP}. Obviously, in this last case there
is no energy ${\cal H}$ which drives the system to thermal
equilibrium. The term $\mu$ in eq.(\ref{eq2}) is a Lagrange multiplier
which ensures that the spherical constraint is satisfied at all times
and the noise $\eta$ satisfies the fluctuation-dissipation relation
$\langle\eta_i(t)\eta_j(s)\rangle=2T \delta(t-s)\delta_{ij}$ where
$\langle...\rangle$ denotes the noise average.

We define the overlap between two configurations of the spins
$\s,\tau$ by the relation $Q=\frac{1}{N}\sum_{i=1}^N\s_i\tau_i$ and
the Hamming distance between these two configurations as,

\be
D=\frac{1-Q}{2}
\label{eq3}
\ee

in such a way that identical configurations have zero distance and
opposite configurations have maximal distance. Now we consider two
copies of the system $\lbrace\s_i,\tau_i\rbrace$ which evolve under the
same noise (\ref{eq1}) but with different initial conditions. Here we
restrict to random initial configurations (i.e equilibrium
configurations at infinite temperature) with initial overlap
$Q(0)$. Nevertheless, different type of initial conditions can be
considered. The different set of correlation functions which describe
the dynamics of the system are given by,

\bea
C(t,s)=\sum_{i=1}^N\s_i(t)\s_i(s)=\sum_{i=1}^N\tau_i(t)\tau_i(s)
\label{eq4a}\\
R(t,s)=\sum_{i=1}^N\frac{\partial\s_i}{\partial h^{\s}_i}=
\sum_{i=1}^N\frac{\partial\tau_i}{\partial h^{\tau}_i}\label{eq4b}\\
Q(t,s)=\sum_{i=1}^N\s_i(t)\tau_i(s)\label{eq4c}
\eea

where $h^{\s}_i,h^{\tau}_i$ are fields coupled to the spins
$\s_i,\tau_i$ respectively. In what follows we take the convention
$t>s$. The previous correlation functions satisfy the boundary
conditions, $C(t,t)=1, R(s,t)=0, \lim_{t\to (s)^+} R(t,s)=1$ while the
two replica overlap $Q(t,s)$ defines the equal time overlap
$Q_d(t)=Q(t,t)$ which yields the Hamming distance at equal times or
damage $D(t)$ through the relation (\ref{eq3}). Following standard
functional methods \cite{BARRAT,CUKU} it is possible to write a closed
set of equations for the previous correlation functions,

\bea
\frac{\partial C(t,s)}{\partial t}+\mu(t)C(t,s)=
\frac{p}{2}\int_0^s du R(s,u)C^{p-1}(t,u)+\nonumber\\
\frac{p(p-1)}{2}
\int_0^t du R(t,u)C(s,u)C^{p-2}(t,u)\label{eq5a}\\
\frac{\partial R(t,s)}{\partial t}+\mu(t)R(t,s)=\delta(t-s)
+\nonumber\\
\frac{p(p-1)}{2}
\int_s^t du R(t,u)R(u,s)C^{p-2}(t,u)\label{eq5b}\\
\frac{\partial Q(t,s)}{\partial t}+\mu(t)Q(t,s)=
\frac{p}{2}\int_0^s du R(s,u)Q^{p-1}(t,u)
+\nonumber\\
\frac{p(p-1)}{2}
\int_0^t du R(t,u)Q(u,s)C^{p-2}(t,u)\label{eq5c}
\eea

while the Lagrange multiplier $\mu(t)$ and the diagonal correlation
function $Q_d(t)$ obey the equations,

\bea
\mu(t)=T+\frac{p^2}{2}\int_0^tdu R(t,u)C^{p-1}(t,u)\label{eq6a}\\
\frac{1}{2}\frac{\partial Q_d(t)}{\partial t}+\mu(t)Q_d(t)=T+
\frac{p}{2}\int_0^t du R(t,u)Q^{p-1}(t,u)
\nonumber\\
+\frac{p(p-1)}{2}
\int_0^t du R(t,u)Q(t,u)C^{p-2}(t,u)\label{eq6b}
\eea

This set of equations is quite involved. For the correlation $C$ and
response functions $R$ eqs.(\ref{eq5a},\ref{eq5b},\ref{eq6a}) several
results are known, in particular their behavior in the FDT regime (where
time translational invariance is satisfied and the
fluctuation-dissipation theorem is obeyed) as well as in the aging regime
\cite{CUKU}.

A first glance to equations (\ref{eq5c}),(\ref{eq6b}) reveals that the
overlap $Q(t,s)$ and its diagonal part $Q_d(t)$ are coupled each
other through the correlation $C(t,s)$ and response function $R(t,s)$.
The trivial solution $Q(t,s)=C(t,s)$ and $Q_d(t)=1$ corresponds to the
case where the initial conditions are the same, $Q_d(0)=1$ and the
distance $D(t)=0$ for all times. This solution (hereafter we will denote
it by HT) is asymptotically reached by the dynamics for high enough
temperatures. The typical time needed to reach that solution grows if 
temperature decreases. At a given temperature (which we
identify with $T_0$) there is an instability in the dynamical equations
(\ref{eq5c}),(\ref{eq6b}) and the asymptotic solution differs from the
HT one. We did not succeed in finding an explicit expression for
$T_0$ but we have been able to show its existence and find
lower and upper bounds for its value.

To show the existence of $T_0$ we focus in the high
temperature FDT regime $(t-s)/t<<1$ with $t,s$ both large and
$TR(t,s)=TR(t-s)=\frac{\partial C(t-s)}{\partial s}$. Writing
$Q(t,s)=Q_d(s)\hat{Q}(t,s)$, using the inequalities $\hat{Q}(t,s)\le
C(t-s)$, $\frac{\partial Q_d(t)}{\partial t}\ge 0$ and inserting these
results in (\ref{eq5c}) it is possible to get the following inequality,

\be T(1-Q_d(t))+\frac{\beta Q_d(t)}{2}(Q_d^{p-2}(t)-1)\ge
\frac{\partial Q_d(t)}{\partial t}\ge 0 
\label{eqineq}
\ee

\noindent A trivial solution which always satisfies that inequality is
$Q_d(t)=1$. It is not difficult to check that the
previous inequality yields a lower bound for the temperature at which
there is an instability in the condition (\ref{eqineq}), leading to

\be
T_0\ge \sqrt{\frac{(p-2)}{2}}~~~~~~~.
\label{eqlower}
\ee

\noindent On the other hand, a linear stability analysis of the equations
(\ref{eq5c}),(\ref{eq6b}) around the HT solution, $Q_d(t)=1-\eps f(t)$, 
$Q(t,s)=C(t-s)-\eps g(t,s)$ where $f(0)=1, g(t,t)=f(t)$ yields for
equation (\ref{eq6b}) in the large $t$ limit,

\be
\frac{1}{2}\frac{\partial f}{\partial t}=-(T-\frac{p\beta}{2})f-
\beta p \int_0^tdu C^{p-1}(t-u)\frac{\partial g(t,u)}{\partial u}~~~.
\label{eqf}
\ee

\noindent Finally, the inequality $\frac{\partial g(t,u)}{\partial
u}\ge 0$ together with (\ref{eqlower}) yields,

\be
\sqrt{\frac{p}{2}-1} \le T_0\le \sqrt{\frac{p}{2}}
\label{equpper}
\ee

As said before, it is very difficult to find an explicit expression for
$T_0$. The reason is that both $Q_d(t)$ and $\hat{Q}(t,s)$ (or
equivalently, $f(t)$ and $g(t,s)$ in the linear stability analysis) are
not related by any $FDT$-like relation in the long time
limit. Consequently, the analysis of the dynamic instability turns out
to be more difficult.

Note that for the particular case $p=2$ the inequality (\ref{equpper})
yields $T_{0}\le 1$. Taking into account that (\ref{equpper}) was
derived under the assumption $T_{0}\ge T_s=1$ that yields $T_{0}=1$. 
The simpler case $p=2$ has been already considered by Stariolo \cite{STA}

In the general case $p>3$ a dynamical instability appears at
temperatures well above any relevant thermodynamic temperature. In
particular, for $p=3$, numerical integration of the dynamical equations
as well as the use of series expansions (see below) yields a dynamical
transition at $T_0(p=3)=1.04\pm 02$ in agreement with the inequalities
(\ref{equpper}). Note that $T_0$ is much higher than $T_d(p=3)=0.6125$ or
$T_s(p=3)=0.5$. The relaxation time $\tau_{relax}$ associated to the decay of
the distance $D(t)$ to zero diverges according to a power
law $\tau_{relax}\simeq (T-T_0)^{-\gamma}$ with $\gamma\simeq 1.1\pm
0.1$.

It is important to note that $T_0$ is not related to any
thermodynamic singularity. In the large $p$ limit the inequality
(\ref{equpper}) yields $T_0\to \sqrt{\frac{p}{2}}$ which gives a
temperature much above the TAP temperature where an exponentially large
number of states start to appear ($T_{TAP}\to
\sqrt{\log(p)}$). Consequently, the origin of the damage spreading
transition is purely dynamical and not related to any thermodynamic
singularity or even to the existence of an exponentially large number of
metastable states in the system.

Now we discuss the behavior of the asymptotic distance below $T_0$. In
principle a new transition at $T_d$ (which we identify as $T_1$) is
expected for the dynamical behavior of $Q_d(t)$ because the correlation
$C$ develops the mode-coupling instability.  We will see that for
$p>2$ the transition temperature $T_1$ is absent. It is very difficult
to get analytical results below $T_d$. A possible way to investigate
the asymptotic long-time limit of $Q_d$ in the low-temperature regime
(i.e below $T_0$) is to numerically integrate the set of dynamical
equations (\ref{eq5a})-(\ref{eq6b}). Unfortunately, the CPU time and
the memory needed to numerically integrate them grows very fast with
the maximum time $t$ (approximately like $t^2$).  On the other hand,
$Q_d(t)$ displays in several cases a non monotonic behavior as a
function of time. Consequently, it is very difficult to extrapolate
the numerical data to the infinite time limit.

An alternative method was recently proposed in \cite{FMP} were the series
expansion for correlation and response functions was used to investigate
the asymptotic long time limit of quantities such as the internal
energy. Here we follow \cite{FMP} but extend their method within the
constrained formalism to include the series expansions for the 
correlation function $Q(t,s)$. To this end, we decompose in Taylor
series the correlation, the response as well as the overlap,

\bea
C(t,s)=\sum_{k=0}^{\infty}\Bigl (\sum_{l=0}^k c_{kl} (t-s)^l t^{k-l}\Bigr )
\label{eq8a}\\
R(t,s)=\sum_{k=0}^{\infty}\Bigl (\sum_{l=0}^k r_{kl} (t-s)^l t^{k-l}\Bigr )
\label{eq8b}\\
Q(t,s)=\sum_{k=0}^{\infty}\Bigl (\sum_{l=0}^k q_{kl} (t-s)^l t^{k-l}\Bigr )
\label{eq8c}\\
\mu(t)=\sum_{k=0}^{\infty}\mu_{k}t^k
\label{eq8d}
\eea

where $c_{k0}=r_{k0}=\delta_{k0}$ and
$Q_d(t)=\sum_{k=0}^{\infty}q_{k0}t^k$. It is possible to write in this
case the recurrence relations between the different coefficients
$c_{kl}, r_{kl}, q_{kl}, \mu_k$. The time necessary to calculate the
first coefficients of the series is not very large and takes a few hours
in a work station to reach the first 80 terms of the series
\footnote{This is true for case $p=3$ while for larger values of $p$
the computational effort is larger}. The radius of convergence of
these series is quite small. To enlarge their radius of convergence we
have used Pade approximants to get an estimate of the asymptotic value
of the distance $Q_d(\infty)$. While the method works very well in
case of the asymptotic value of the energy \cite{FMP} (which depends
on the Lagrange multiplier $\mu$ via the relation $\mu=T-pE(t)$) it is
less effective for the asymptotic distance. The reason is that, while
$\mu(t)$ is a monotonous increasing function of time, $Q_d(t)$ is
not. In fact, below $T_0$ the overlap $Q_d(t)$ has a maximum as a
function of time for some values of $T$ and the initial condition
$Q_d(0)$. Consequently, the complex function $Q_d(z)$ turns out to
have zeros close to the real axis and the radius of convergence of the
Pades is smaller. But still it is possible to obtain some estimates
for the asymptotic distance. Numerical integrations of the dynamical
equations have been used to check that our extrapolations in the
infinite time limit are correct.

\begin{figure}
\begin{center}
\leavevmode
\epsfysize=230pt{\epsffile{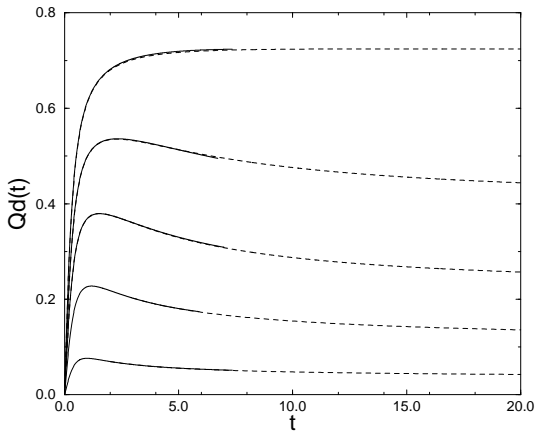}}
\end{center}
  \protect\caption[1]{$Q_d(t)$ with $Q_d(0)=0$ as a function of time
  for different temperatures. From top to bottom
  $T=0.9,0.7,0.5,0.3,0.1$. The continuous lines are the numerical
  integrations with time step $\Delta t=0.01$ and the dashed lines are
  the reconstructed functions obtained from the Pade analysis.
  \protect\label{FIG1} }
\end{figure}

Some of our results are shown in figures 1 and 2 for case $p=3$. We
have studied three different initial conditions: a) anti-correlated
random initial conditions with $Q_d(0)=-1$, b) uncorrelated random
initial conditions with $Q_d(0)=0$ and c) partially correlated random
initial conditions with $Q_d(0)=0.5$. Case a) was analyzed using
diagonal and the first off-diagonal Pade approximants assuming an
asymptotic power law decay $Q_d(t)=Q_d(\infty)+At^{-\gamma}$.  Cases
b) and c) turned out to be more difficult to analyze due to the small
radius of convergence of the series as well as to the presence of
poles in the Pades.

The behavior of $Q_d(t)$ for case b) is shown in figure 1 for
different temperatures. The continuous line corresponds to the
numerical integration of the dynamical equations while the dashed
lines are the reconstructed functions $Q_d(t)$ obtained from the Pade
analysis. Note the presence of a maximum in $Q_d(t)$ for several
different temperatures. This feature is a consequence of the
non-linear character (in $Q$) of equations (\ref{eq5c},\ref{eq6b}) for
$p\ge 3$ and is absent for $p=2$ \cite{STA}.

Figure 2 shows the asymptotic distance $D_{\infty}$ for cases a),b)
and c) as a function of the temperature. We find that the asymptotic
distance is independent of the initial correlation. This is an
interesting result since one would expect (at least below $T_d$) a
dependence on the initial condition.  We have to note that the
dependence on the initial conditions is expected for models with the
symmetry $\s\to -\s$ and with a simple free energy landscape. If the
initial distance $D_0$ is not zero there is always a finite
probability (depending on $D_0$) that the initial configurations start
in different (symmetry related) ergodic components. Consequently, the
asymptotic distance depends on the initial value of $D_0$. Here, such
behavior is only found for $p=2$ \cite{STA}.

\begin{figure}
\begin{center}
\leavevmode
\epsfysize=230pt{\epsffile{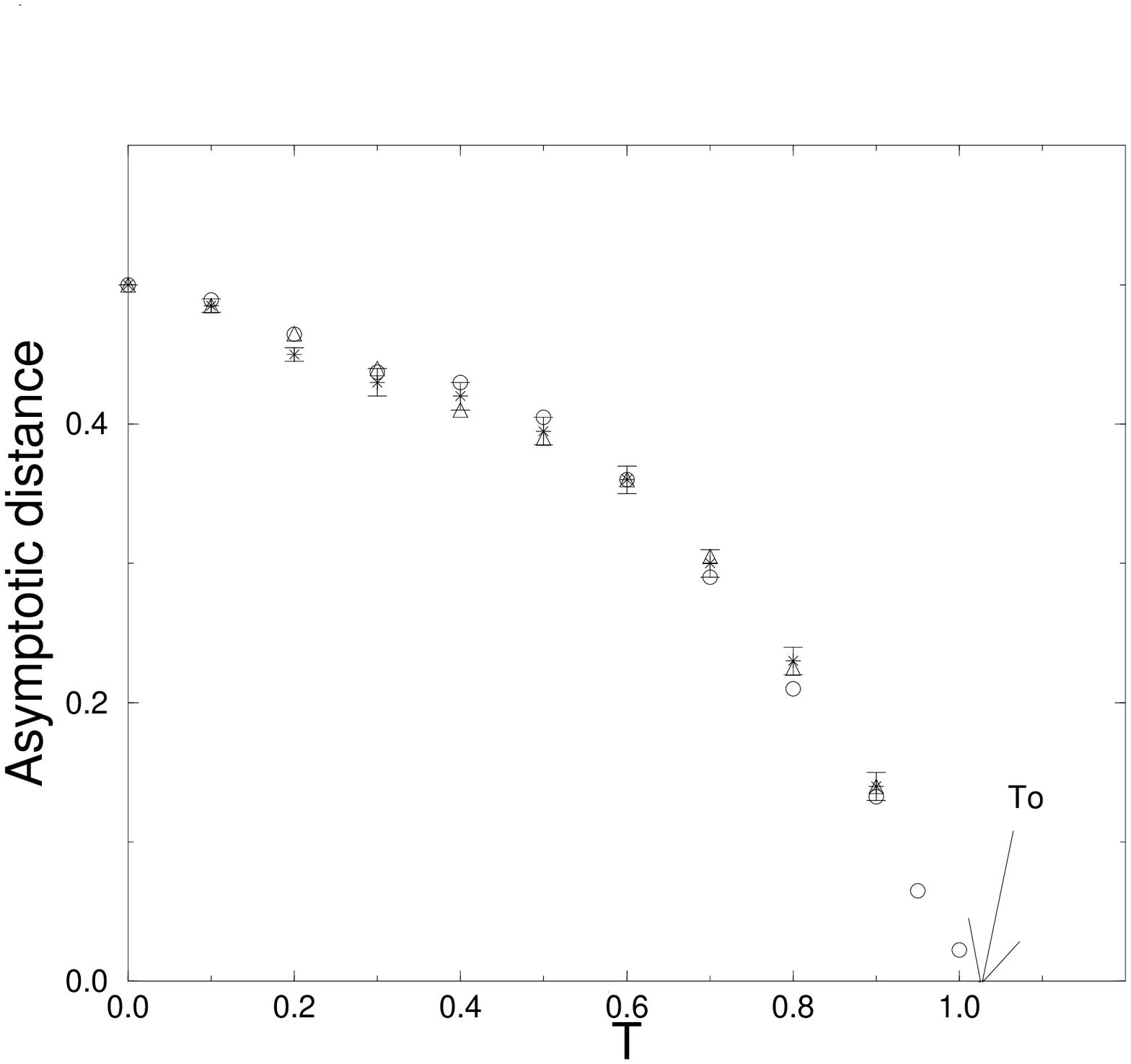}}
\end{center}
  \protect\caption[2]{Asymptotic distance $D_{\infty}$ for $p=3$
obtained from the Pade analysis of the series expansions for different
initial conditions $D_0=1$ (circles), $D_0=0.5$ (triangles),
$D_0=0.25$ (stars). Typical error bars are shown for the last case.
\protect\label{FIG2} }
\end{figure}


In summary, we have studied the spreading of damage in the
off-equilibrium mode-coupling equations. We have explicitly shown the
existence of a damage spreading transition $T_0$ and also found lower
and upper bounds for its value. This transition takes place at very
high temperatures. On the other hand, this transition is completely
unrelated to the existence of metastable states in the system. In
fact, we have observed that this transition is quite stable to the
inclusion of any degree of asymmetry. Indeed in the fully asymmetric
case we find that $T_0\simeq 0.71$ for $p=3$. Consequently, the damage
spreading transition persists in the absence of the spin-glass phase or
even in the absence of metastable states. This result corroborates
some results already found for other disordered spin-glass
models \cite{DE,ABC}.  Interestingly, equations
(\ref{eq5c},\ref{eq6b}) show that, only for $p>2$, the coupling
between the damage $Q_d(t)$ and the two-time correlation function
$Q(t,s)$ is non linear. This non-linear coupling is crucial for the
appearance of the damage transition $T_0$ which is well above
$T_d$. We have also shown that below $T_0$ the asymptotic distance is
independent of the initial distance. This was unexpected since such a
dependence has been usually found in numerical investigations of
several spin-glass models \cite{DW}.  It would be interesting to
understand whether this result is a direct consequence of the first
order nature of the glass transition. This and other issues, such
as the scaling behavior of the overlap $Q(t,s)$ in the aging regime
and the existence of this transition in non-disordered glass forming
liquids are left for future investigations.

{\bf Acknowledgments}.  We acknowledge stimulating discussions with
Th. M. Nieuwenhuizen and W. A. Van Leeuwen. The work by F.R has been
supported by FOM under contract FOM-67596 (The Netherlands).

\hspace{-2cm}

\vfill

\end{document}